\documentclass[letterpaper, 10 pt, conference]{ieeeconf}
\IEEEoverridecommandlockouts                              
\usepackage{slashbox}
\usepackage{amssymb}
\usepackage{amsmath}
\usepackage{graphicx}
\usepackage{slashbox}
\usepackage{array}
\usepackage{framed}
\usepackage{color}
\usepackage{graphicx}          
\usepackage{balance}
\usepackage{amsfonts}

\newtheorem{Theorem 1}{Theorem}
\newtheorem{Theorem 2}[Theorem 1]{Theorem}
\newtheorem{Theorem 3}[Theorem 1]{Theorem}
\newtheorem{Theorem 4}[Theorem 1]{Theorem}
\newtheorem{Theorem 5}[Theorem 1]{Theorem}
\newtheorem{Theorem 6}[Theorem 1]{Theorem}
\newtheorem{Theorem 7}[Theorem 1]{Theorem}
\newtheorem{Theorem 8}[Theorem 1]{Theorem}
\newtheorem{Assumption 1}{Assumption}
\newtheorem{Assumption 2}[Assumption 1]{Assumption}
\newtheorem{Assumption 3}[Assumption 1]{Assumption}
\newtheorem{Assumption 4}[Assumption 1]{Assumption}
\newtheorem{Assumption 5}[Assumption 1]{Assumption}
\newtheorem{Remark 1}{Remark}
\newtheorem{Remark 2}[Remark 1]{Remark}
\newtheorem{Remark 3}[Remark 1]{Remark}
\newtheorem{Remark 4}[Remark 1]{Remark}
\newtheorem{Remark 5}[Remark 1]{Remark}
\newtheorem{Remark 6}[Remark 1]{Remark}
\newtheorem{Remark 7}[Remark 1]{Remark}
\newtheorem{Remark 8}[Remark 1]{Remark}
\newtheorem{Remark 9}[Remark 1]{Remark}
\newtheorem{Remark 10}[Remark 1]{Remark}
\newtheorem{Remark 11}[Remark 1]{Remark}
\newtheorem{Lemma 1}{Lemma}
\newtheorem{Lemma 2}[Lemma 1]{Lemma}
\newtheorem{Lemma 3}[Lemma 1]{Lemma}
\newtheorem{Definition 1}{Definition}
\newtheorem{Definition 2}[Definition 1]{Definition}
\newtheorem{Definition 3}[Definition 1]{Definition}
\newtheorem{Definition 4}[Definition 1]{Definition}
\newtheorem{Definition 5}[Definition 1]{Definition}
\newtheorem{Problem 1}{Problem}

\begin{document}
\title{\LARGE \bf The collective oscillation period of inter-coupled Goodwin oscillators}
\author{Yongqiang Wang, Yutaka Hori, Shinji Hara {\it Fellow, IEEE}, Francis J. Doyle III {\it Fellow, IEEE}\thanks{The work was supported in part by Institute for Collaborative Biotechnologies under
 grant W911NF-09-D-0001, U.S. Army
Research Office under grant W911NF-07-1-0279,  National Institutes
of Health under grant GM078993, and  Grant-in-Aid for JSPS Fellows
 of Japan Society for the Promotion of Science (JSPS) under grant
No. 23-9203. The content of the information does not necessarily
reflect the position or the policy of the Government, and no
official endorsement should be inferred.}
\thanks{Yongqiang Wang,  Francis J. Doyle III are with Department of Chemical Engineering, University of California, Santa
Barbara,  California 93106-5080 USA. E-mail: wyqthu@gmail.com,
frank.doyle@icb.ucsb.edu. }\thanks{ Yutaka Hori and Shinji Hara are
with Department of Information Physics and Computing, The University
of Tokyo, Tokyo 113-8656 Japan. E-mail:
Yutaka\_hori@ipc.i.u-tokyo.ac.jp, Shinji\_hara@ipc.i.u-tokyo.ac.jp.
}
} \maketitle      

\begin{abstract}
Many biological oscillators are arranged in networks composed of
many inter-coupled cellular oscillators. However, results are still
lacking on the collective oscillation period of inter-coupled gene
regulatory oscillators, which, as has been reported, may be
different from the oscillation period of an autonomous cellular
oscillator. Based on the Goodwin oscillator, we analyze the
collective oscillation pattern of coupled cellular oscillator
networks. First we give a condition under which the oscillator
network exhibits oscillatory and synchronized behavior, then we
estimate the collective oscillation period based on a multivariable
harmonic balance technique. Analytical results are derived in terms
of biochemical parameters, thus giving insight into the basic
mechanism of biological oscillation and providing guidance in
synthetic biology design. Simulation results are given to confirm
the theoretical predictions.
\end{abstract}
%

\section{Introduction}

Diverse biological rhythms are generated by multiple cellular
oscillators that somehow manage to operate synchronously. In systems
ranging from circadian rhythms to segmentation clocks, it remains a
 challenge to understand how collective oscillation patterns
(e.g., oscillation period, amplitude) arise from
 autonomous cellular oscillators. As has been reported in the
 literature, there can be significant differences between collective oscillation patterns and
cell autonomous oscillation patterns, the difference are embodied
not only in the oscillation amplitude \cite{popovych:11}, but also
in the oscillation period  \cite{wunsche:05,Herrgen:10}.

 The famous Goodwin
oscillator provides a perfect model to study the mechanism of how
collective oscillation pattern arises from autonomous cellular
oscillators. The Goodwin oscillator was proposed in 1965 to model
the oscillatory behavior in enzymatic control processes
\cite{Goodwin:65, Goodwin:66}. It is a minimal model that describes
the oscillatory negative feedback regulation of a translated protein
which inhibits its own transcription. Because the Goodwin oscillator
can capture the essential characteristics in biochemical
oscillators, it has been extensively used to model biological
oscillators such as ultradian clocks of vertebrate embryos
\cite{Zeiser:07} and circadian clocks of neurospora \cite{Ruoff:01},
drosophila \cite{Ruoff:99} as well as mammals \cite{Gonze:05}.

Another advantage of the Goodwin oscillator is that it allows for an
analytical understanding of the basic dynamical mechanisms, whereas
other biophysically substantiated models of cellular oscillators
usually exhibit high complexity and large number of variables, which
hamper the mathematical treatment, and subsequently obscure the
underlying mechanisms. The Goodwin oscillator allows one to gain
insights into the mechanisms of biochemical rhythms. For example,
the oscillation conditions of a single Goodwin oscillator were
obtained in \cite{Griffith:68, Hunding:74,Tyson:75,Hori:11}. The
synchronization conditions for coupled Goodwin oscillator networks
were reported in \cite{stan_acc:07,Kim:07,Hamadeh:12}. Based on a
multivariable harmonic balance technique, the
 oscillation patterns of a single Goodwin oscillator were obtained in
 \cite{Rapp:76, Hori:10}. This is an important step toward
 understanding the period determination in biochemical oscillators.
 However, given that biological rhythms are generated by \textbf{multiple} cellular
 oscillators coupled through intercellular signaling,
   it remains a challenge to determine the periods
 in biological rhythms, which ranges from seconds in cardiac
 cell contraction  to years in
 reproduction. Recently, using the phenomenological phase
 model, the authors in \cite{Liu:97} proved that if the intercellular coupling is weak,  the collective
 period is identical to the autonomous period. However, since the phase model contains no direct biological
mechanism of the cellular clock, it can potentially weaken the
model's reliability in checking scientific hypotheses.

 This paper derives analytical results for the collective
 oscillation period of multi-cellular networks based on coupled Goodwin
 oscillators. Specifically, we study the collective  period of a
 network of Goodwin oscillators connected by diffusive
 coupling. The basic idea is to use a multivariable harmonic
 balance technique \cite{Iwasaki:08}. In fact, due to the multi-cellular structure,
the solution to harmonic balance equations in the multivariable
harmonic balance technique is very difficult to obtain. Here we are
interested in the collective period, so we can circumvent the
problem by restricting our attention to solutions corresponding to
synchronized oscillations in the oscillator network. To this end, we
also give an oscillation/synchronization condition for the Goodwin
oscillator network.

\section{Model description and transformation}

\begin{figure}[!hbp]
\begin{center}
  \includegraphics[width=0.5\columnwidth]{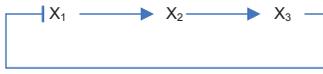}
    \caption{{\bf Schematic diagram of a Goodwin oscillator.} The mRNA $\textmd{X}_1$
produces a protein $\textmd{X}_2$ which, in turn, activates a
transcriptional inhibitor $\textmd{X}_3$. The inhibitor
$\textmd{X}_3$ inhibits the production of $\textmd{X}_1$, which
closes a negative feedback loop.}
    \label{fg:goodwin oscillator}
\end{center}
\end{figure}

 The Goodwin oscillator describes the dynamics of an
oscillatory negative feedback regulation loop. As  shown in Fig.
\ref{fg:goodwin oscillator},   mRNA $\textmd{X}_1$ produces a
protein $\textmd{X}_2$ which, in turn, activates a transcriptional
inhibitor $\textmd{X}_3$. The inhibitor $\textmd{X}_3$ inhibits the
production of mRNA $\textmd{X}_1$, which closes a negative feedback
loop. The kinetic dynamics of the Goodwin oscillator are given by
\cite{Goodwin:65, Griffith:68,Fall:05}:
\begin{equation}\label{eq:original Goodwin model}
\left\{
\begin{aligned}
\frac{d[\textmd{X}_1]}{dT}&=\frac{v_0}{1+\left([\textmd{X}_3/K_m]^p\right)}-k_1[\textmd{X}_1]\\
\frac{d[\textmd{X}_2]}{dT}&=v_1[\textmd{X}_1]-k_2[\textmd{X}_2]\\
\frac{d[\textmd{X}_3]}{dT}&=v_2[\textmd{X}_2]-k_3[\textmd{X}_3]
\end{aligned}
\right.
\end{equation}
Here $[\textmd{X}_1],$ $[\textmd{X}_2]$, and $[\textmd{X}_3]$ are
concentrations of mRNA $\textmd{X}_1$, protein $\textmd{X}_2$, and
inhibitor $\textmd{X}_3$, respectively; $v_0$, $v_1$, and $v_2$ are
the rates of transcription, translation, and catalysis; $k_1$,
$k_2$, and $k_3$ are rate constants for degradation of each
component; $1/K_m$ is the binding constant of end product to
transcription factor; and $p$ is Hill coefficient, which describes
cooperativity of end product repression.

There are other interpretations of the above Goodwin oscillator
 \cite{Goodwin:65}, for example, one
may regard $\textmd{X}_2$ as an enzyme precursor that after primary
synthesis on mRNA templates, $\textmd{X}_1$, passes through a pool
of inactive molecules before being transformed into mature, active
enzyme, $\textmd{X}_3$. One may also take $\textmd{X}_1$ to be an
enzyme population whose rate of synthesis is regulated by feedback
control at the polysome level via a metabolite $\textmd{X}_3$. In
this case, $\textmd{X}_2$ is then an intermediate in the
biosynthetic sequence leading to $\textmd{X}_3$.

The Goodwin oscillator (\ref{eq:original Goodwin model}) can be
transformed into

\begin{equation}\label{eq:Goodwin oscillator}
\left\{
\begin{aligned}
\frac{dx_1}{dt}&= f(x_3)-b_1x_1\\
\frac{dx_2}{dt}&=x_1-b_2x_2 \\
\frac{dx_3}{dt}&=x_2-b_3x_3
\end{aligned}
\right., \quad f(x)= \frac{1}{1+x^p}
\end{equation}
via the introduction of dimensionless variables
\[
\varsigma=\sqrt[3]{\frac{K_m}{v_0v_1v_2}},\:
x_1=\frac{\varsigma^2v_1v_2[\textmd{X}_1]}{K_m},
\]
\[
x_2=\frac{\varsigma v_2[\textmd{X}_2]}{K_m},\:
x_3=\frac{[\textmd{X}_3]}{K_m},\: t=\frac{1}{\varsigma} T
\]
In (\ref{eq:Goodwin oscillator}), $b_1$, $b_2$, and $b_3$ are
positive parameters given by
\[
b_i=k_i\varsigma,\quad i=1,2,3
\]

%

Suppose the oscillator network is composed of $N$ oscillators, and
they are connected by diffusive coupling, then the dynamics of the
network are given by
\begin{equation}\label{eq:Goodwin oscillator_entrain_network}
\left\{
\begin{aligned}
\frac{dx_{1,i}}{dt}=&f(x_{3,i})-b_1x_{1,i}\\
\frac{dx_{2,i}}{dt}=&x_{1,i}-b_2x_{2,i}-\sum\limits_{j=1}^{N}a_{i,j}(x_{2,i}-x_{2,j})\\
\frac{dx_{3,i}}{dt}=&x_{2,i}-b_3x_{3,i}
\end{aligned}
\right.
\end{equation}
where $i=1,2,\ldots,N$ denotes the index of the $i$th oscillator,
and $a_{i,j}\geq 0$  denotes the coupling strength  between
oscillator $i$ and oscillator $j$. If $a_{i,j}=0$, then there is no
interaction between oscillator $i$  and oscillator $j$.

\begin{Assumption 1}\label{as:assumption connected}
We assume that the interaction is bidirectional, i.e.,
$a_{i,j}=a_{j,i}$. We also assume that the interaction topology is
connected, i.e., there is a multi-hop path (i.e., a sequence with
nonzero values
$a_{i,m_1},\,a_{m_1,m_2},\,\ldots,\,a_{m_{p-1},m_{p}},\,a_{m_p,j}$)
from each node $i$ to every other node $j$.
\end{Assumption 1}
\begin{Remark 1}
Since protein is usually diffusible and involved in intercellular
signaling \cite{Lauffenburger:93}, we model the intercellular
coupling  by the diffusion of $x_2$, which usually represents the
protein product in the Goodwin oscillator model.
\end{Remark 1}

For convenience in analysis, we can write (\ref{eq:Goodwin
oscillator_entrain_network}) into the following matrix form:
\begin{equation}\label{eq:matrix form}
\left\{
\begin{aligned}
\frac{dX_{1}}{dt}=&\vec{f}(X_3)-b_1X_1\\
\frac{dX_{2}}{dt}=&X_{1}-b_2X_{2}-AX_2\\
\frac{dX_{3}}{dt}=&X_{2}-b_3X_{3}
\end{aligned}
\right.
\end{equation}
where
\begin{equation}\label{eq:f(X_3)}
X_i=\left[\begin{array}{c}x_{i,1}\\x_{i,2}\\\vdots\\x_{i,N}\end{array}\right],\:
i=\{1,2,3\},\: \vec{f}(X_3)=\left[\begin{array}{c} f(x_{3,1})\\
f(x_{3,2})\\\vdots\\f(x_{3,N})\end{array}\right],\:
\end{equation}
\begin{equation}\label{eq:A}
\begin{aligned}
&A=\\
&\left[\begin{array}{cccc}\hspace{-0.05cm} \sum_{j\neq
1}a_{1,j}&-a_{1,2}&\ldots&-a_{1,N}\\-a_{2,1}
&\hspace{-0.05cm}\sum_{j\neq
2}a_{2,j}&\ldots&-a_{2,N}\\\vdots&\ddots&\ddots&\vdots\\-a_{N,1}&\ldots&-a_{N,N-1}&\hspace{-0.05cm}\sum_{j\neq
N}a_{N,j}\end{array}\right]
\end{aligned}
\end{equation}

Next, based on a multivariable harmonic balance technique, we will
analyze the collective period of the Goodwin oscillator
network in (\ref{eq:matrix form}). 


\section{Oscillation/synchronization condition}

To study the collective period, we need to guarantee that the
component elements in $X_i$ $(i=\{1,2,3\})$ in (\ref{eq:matrix
form})  first oscillate, and then further more, oscillate in
synchrony. In this paper, we consider the Y-oscillation, which is
defined as follows \cite{Iwasaki:08}:
\begin{Definition 4}
A system $\dot{x}=f(x)$ with $x(t)\in\mathcal{R}^m$ is said to be
Y-oscillatory if each solution is bounded and there exists a state
variable $x_i$ such that
\begin{math}\lim\limits_{\overline{_{t\hspace{-0.02cm}\rightarrow\hspace{-0.03cm}+\infty }}}x_i(t)<\overline{\lim\limits_{_{t\hspace{-0.02cm}\rightarrow\hspace{-0.05cm}+\infty}}}x_i(t)\end{math}
for almost all initial states $x(0)$.
\end{Definition 4}

To prove that (\ref{eq:matrix form}) is Y-oscillatory,  we introduce
Lemma \ref{le:lemma 1}:
\begin{Lemma 1}\label{le:lemma 1}\cite{Pogromsky:99}
System (\ref{eq:matrix form}) is Y-oscillatory if all conditions
(a), (b), and (c) hold:
\begin{itemize}
\item[(a)] It only has isolated equilibria $X^{\ast}$;
\item[(b)] The positive semiorbit $\{X(t)\triangleq[X_1(t),X_2(t),X_3(t)]\big|t\geq 0 \:\textrm{and}\: t\in\textrm{dom}
X(\bullet)\}$ is bounded;
\item[(c)] The Jacobian matrix  evaluated at the equilibrium point
$X^{\ast}$, i.e., $J$, has at least one eigenvalue with positive
real part.
\end{itemize}
\end{Lemma 1}
\begin{proof}
The Lemma can be obtained by combining Theorem 1 in
\cite{Pogromsky:99} and the discussion below its proof which shows
that the hyperbolicity condition can be relaxed.
\end{proof}

In the following, we will prove that (\ref{eq:matrix form})
satisfies all the conditions in Lemma  \ref{le:lemma 1}, and hence
is Y-oscillatory. We will also give a synchronization condition for
the oscillation. In this manuscript, we define synchronization as
follows:
\begin{Definition 5}\label{de:synchrony}
System (\ref{eq:matrix form}) is synchronized if the elements in
$X_3$ satisfy
$\lim\limits_{t\rightarrow+\infty}|x_{3,i}(t)-x_{3,j}(t)|=0$ for any
$1\leq i,j\leq N$.
\end{Definition 5}
\begin{Remark 1}
Only $x_{3,i}$ ($i=1,2,\ldots,N$) is used in the definition of
synchronization. This is because according to the modeling
assumption, $x_{3,i}$ is corresponding to the concentration of
inhibitor or enzyme, which can be regarded as the output of an
Goodwin oscillator. Note that under this definition, when the system
is synchronized, $x_{1,i}$ ($i=1,2,\ldots,N$) may be identical or
non-identical. The same situation holds for $x_{2,i}$
($i=1,2,\ldots,N$).
\end{Remark 1}
\begin{Theorem 1}\label{th:theorem 1}
(\ref{eq:matrix form}) has oscillatory  solutions if it satisfies
the following inequality
\begin{equation}\label{eq:condition}
 R\triangleq\frac{pb_1^2b_2^2b_3^2x_0^{p+1}} { (b_1+b_2+b_3)(b_1b_2+b_2b_3+b_1b_3)-b_1b_2b_3}> 1
\end{equation}
where $x_0$ is the  unique  positive solution to $
1/(1+x_0^p)=b_1b_2b_3x_0.$ Furthermore,  the oscillations in all
oscillators are synchronized if the algebraic connectivity $\varrho$
(which is defined as the second smallest eigenvalue of matrix $A$ in
(\ref{eq:A})) of coupling topology satisfies
\begin{equation}\label{eq:synchornization condition}
\varrho > -b_1+\frac{\gamma}{4b_2b_3},\:
\gamma=\max\limits_{x\geq0}\left\{\frac{px^{p-1}}{(1+x^p)^2}\right\}
\end{equation}
\end{Theorem 1}

\begin{proof}
From Lemma \ref{le:lemma 1}, we know to guarantee a Y-oscillatory
solution, we need to prove the conditions
 (a), (b), and (c), are satisfied.
The proof is decomposed into three steps.

\subsubsection*{Step I $-$ Satisfaction of condition (a)}

Using the monotonic property of $f(\bullet)$, we can prove that
(\ref{eq:Goodwin oscillator_entrain_network}) has only one
equilibrium point
$X_3^{\ast}=\left[\begin{array}{cccc}x_0&x_0&\ldots&x_0\end{array}\right]^T$
with $x_0>0$ determined by $f(x_0)=b_1b_2b_3x_0$. The derivation is
as follows:

 In the equilibrium point,
we have
\[
\frac{dX_i^{\ast}}{dt}=0,\:i=\{1,2,3\}
\]
So after some simple algebra, (\ref{eq:matrix form}) is reduced to
\begin{equation}\label{eq:equilibrium point}
\vec{f}(X_3^{\ast})-b_1b_2b_3X_3^{\ast}=b_1b_3AX_3^{\ast}
\end{equation}
 where $\vec{f}(X_3^{\ast})$ and $A$ are defined in
(\ref{eq:f(X_3)}) and (\ref{eq:A}), respectively. To prove the
uniqueness of the equilibrium point, we only need to prove that
(\ref{eq:equilibrium point}) has a unique solution.

To this end, we check equation (\ref{eq:equilibrium point})
element-wisely. Making use of the structure of matrix $A$ in
(\ref{eq:A}), we have the following equation for all
$i=1,2,\ldots,N$:
\begin{equation}\label{eq:element-wisely}
\begin{aligned}
g({x_{3,i}^{\ast}})\triangleq
f(x_{3,i}^{\ast})-b_1b_2b_3x_{3,i}^{\ast} =b_1b_3\sum_{j\neq
i}a_{i,j}(x_{3,i}^{\ast}-x_{3,j}^{\ast})
\end{aligned}
\end{equation}

Since the interaction is bi-directional, i.e., $a_{i,j}=a_{j,i}$, it
follows
\begin{equation}\label{eq:sum}
\sum_{i=1}^Ng(x_{3,i}^{\ast})=\sum_{i=1}^N\sum_{j\neq
i}a_{i,j}(x_{3,i}^{\ast}-x_{3,j}^{\ast})=0
\end{equation}
Next, we prove that (\ref{eq:equality of g()}) holds
 by proving that both
$\max\limits_i\{g(x_{3,i}^{\ast})\}$ and
$\min\limits_i\{g(x_{3,i}^{\ast})\}$ are zero:
\begin{equation}\label{eq:equality of g()}
g(x_{3,1}^{\ast})=g(x_{3,2}^{\ast})=\ldots=g(x_{3,N}^{\ast})=0
\end{equation}

Suppose to the contrary that (\ref{eq:equality of g()}) does not
hold, and hence  $\max\limits_i\{g(x_{3,i}^{\ast})\}>0$ is satisfied
since $\sum_{i=1}^Ng(x_{3,i}^{\ast})=0$ holds according to
(\ref{eq:sum}). Represent the index that has the largest
$g(x_{3,i}^{\ast})$ among all $i$ ($i=1,2,\ldots,N$) as $m$. If
there are multiple indices $m$ satisfying
$g(x_{3,m}^{\ast})=\max\limits_i\{g(x_{3,i}^{\ast})\}$, then any one
can be index $m$. Because $f(\bullet)$ is a decreasing function, it
follows that $g(\bullet)$ is a decreasing function according to its
definition on the left hand side of (\ref{eq:element-wisely}). So if
the maximal $g(x_{3,i}^{\ast})$ is attained when $i=m$,
$x_{3,m}^{\ast}$ should be the smallest among
$x_{3,1}^{\ast},\,x_{3,2}^{\ast},\ldots,x_{3,N}^{\ast}$. Therefore,
the right hand side of  (\ref{eq:element-wisely}), i.e.,
\begin{equation}\label{eq:maximal}
\begin{aligned}
 b_1b_3\sum_{j\neq i}a_{m,j}(x_{3,m}^{\ast}-x_{3,j}^{\ast})
\end{aligned}
\end{equation}
should be non-positive, and hence $g(x_{3,m}^{\ast})$ should be
non-positive. This contradicts the fact that $g(x_{3,m}^{\ast})$ is
the largest among all $g(x_{3,i}^{\ast})$ and it is positive (due to
the constraint in (\ref{eq:sum})). Hence
$\max\limits_i\{g(x_{3,i}^{\ast})\}=0$ holds.

Similarly, we can prove that $\min\limits_i\{g(x_{3,i}^{\ast})\}=0$
holds.

Therefore, we have (\ref{eq:equality of g()}), which further leads
to
\begin{equation}\label{eq:equality of f()}
f(x_{3,i}^{\ast})=b_1b_2b_3 x_{3,i}^{\ast},\quad i=1,2,\ldots,N
\end{equation}
%
%
Given that $f(\bullet)$ is a monotonic decreasing function on
$\mathbb{R_+}$, it follows that the solution to (\ref{eq:equality of
f()}) is unique and it satisfies
\begin{equation}\label{eq:equality}
x_{3,1}^{\ast}=x_{3,2}^{\ast}=\ldots=x_{3,N}^{\ast}=x_0>0,\quad
f(x_{0})=b_1b_2b_3 x_{0}
\end{equation}
 Therefore  the solution to
(\ref{eq:equilibrium point}) is unique, thus the equilibrium point
 is unique and hence isolated.

\subsubsection*{Step II $-$ Satisfaction
of condition (b)}

 Following the derivations
in \cite{Mallet:96,Samad:05}, we can easily get that condition (b)
is satisfied.

\subsubsection*{Step III $-$ Satisfaction of condition (c)} The fact that $J$ has at least one eigenvalue with positive
real part is equivalent to the statement that the linearized system
of (\ref{eq:matrix form}) around the equilibrium point is strictly
unstable. So instead of proving (c) directly, next we prove the
strict instability of linearized system of (\ref{eq:matrix form})
around the equilibrium point under condition (\ref{eq:condition}).
To this point, we transform (\ref{eq:matrix form}) into the
frequency domain as shown in Fig. \ref{fg:schematic}, where $H(s)$
is given by
\begin{eqnarray}\label{eq:H(s) in theory I}
\begin{aligned}
H(s)&=\big((sI+b_1I)(sI+b_2I+A)(sI+b_3I)\big)^{-1}\\
&=\frac{1}{(s+b_1)(s+b_3)}(sI+b_2I+A)^{-1}
\end{aligned}
\end{eqnarray}

\begin{figure}[!hbp]
\begin{center}
  \includegraphics[width=0.7\columnwidth]{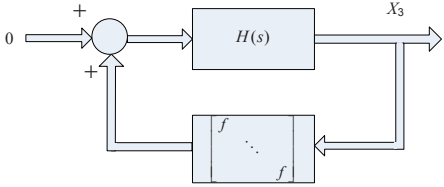}
    \caption{{ Schematic diagram of the frequency domain formulation of (\ref{eq:matrix form}).} }
    \label{fg:schematic}
\end{center}
\end{figure}

 Linearizing the  nonlinear item $\vec{f}(X_3)$ in (\ref{eq:matrix form}) around the
equilibrium point
$X_3^{\ast}=\left[\begin{array}{cccc}x_0&x_0&\ldots&x_0\end{array}\right]^T$
yields
\begin{equation}\label{eq:approximation}
\vec{f}(X_3)=\sigma X_3,
\:\sigma=-\frac{px_0^{p-1}}{(1+x_0^p)^2}=-px_0^{p-1}(b_1b_2b_3x_0)^2
\end{equation}
In the second equality, the relation $f(x_{0})=b_1b_2b_3 x_{0}$ at
the equilibrium point is employed.

Based on the linearization in (\ref{eq:approximation}), we can get
the closed-loop dynamics of the system in Fig. \ref{fg:schematic} as
\begin{equation}\label{eq:G}
G(s)=(I-\sigma H(s))^{-1}H(s)
\end{equation}

Since $A$ is a symmetric matrix, it only has real eigenvalues and it
can always be diagonalized as follows:
\begin{equation}\label{eq:diagonization of A}
A=P\Upsilon P^{-1},\quad
\Upsilon=\textrm{diag}(\upsilon_1,\:\upsilon_2,\ldots\:\upsilon_N)
\end{equation}
where $P$ is the similarity transformation matrix. Since $A$ is the
graph Laplacian, it always has an eigenvalue $0$ associated with  an
eigenvector composed of identical elements \cite{horn:85}. Here we
arrange the eigenvalues in increasing order, so $\upsilon_1=0$
always holds and the connectivity assumption in Assumption
\ref{as:assumption connected} leads to $\upsilon_i>0$
($i=2,3,\ldots,N$) \cite{horn:85}.

Substituting (\ref{eq:diagonization of A}) into (\ref{eq:H(s) in
theory I}), we have
\begin{equation}\label{eq:H(s)}
H(s)=P\Lambda
P^{-1},\:\Lambda=\textrm{diag}(\lambda_1,\:\lambda_2,\ldots\:\lambda_N)
\end{equation}
where eigenvalues $\lambda_i$ $(i=1,2,\ldots,N)$ are given by
\[
\begin{aligned}
\lambda_1&=\frac{1}{(s+b_1)(s+b_2)(s+b_3)}\\
\lambda_j&=\frac{1}{(s+b_1)(s+b_2+\upsilon_j)(s+b_3)},\quad
j=2,3,\ldots,N
\end{aligned}
\]

Using (\ref{eq:G}) and (\ref{eq:H(s)}), we can obtain that
\begin{equation}\label{eq:G(s)}
G(s)=P(I-\sigma\Lambda)^{-1}\Lambda P^{-1}=P\Delta P^{-1}
\end{equation}
where
\[
\Delta=\textrm{diag}(\delta_1,\:\delta_2,\ldots,\:\delta_N)
\]
with eigenvalues $\delta_i$ $(i=1,2,\ldots,N)$ given by
\begin{eqnarray}\label{eq:delta_1 in G(s)}
\delta_1
=\frac{\lambda_1}{1-\sigma\lambda_1}=\frac{1}{(s+b_1)(s+b_2)(s+b_3)-\sigma}\qquad\quad
\end{eqnarray}
\begin{eqnarray}\label{eq:delta_j in G(s)}
\delta_j
=\frac{\lambda_j}{1-\sigma\lambda_j}=\frac{1}{(s+b_1)(s+b_2+\upsilon_j)(s+b_3)-\sigma},
\end{eqnarray}
where $j=2,3,\ldots,N$.


According to the Routh$-$Hurwitz stability criterion, we know
$\delta_1$ is strictly unstable if and only if
\begin{equation}\label{eq:delta_2}
\sigma< b_1b_2b_3-(b_1+b_2+b_3)(b_1b_2+b_1b_3+b_2b_3)
\end{equation}
is satisfied, and $\delta_j$ ($j=2,3,\ldots,N$) is strictly unstable
if and only if
\begin{equation}\label{eq:delta_1}
\begin{aligned}
\sigma<&
b_1(b_2+\upsilon_j)b_3-(b_1+b_2+\upsilon_j+b_3)\times\\
&\qquad\big(b_1(b_2+\upsilon_j)+b_1b_3+(b_2+\upsilon_j)b_3\big)
\end{aligned}
\end{equation}
is satisfied. Note that (\ref{eq:delta_2}) is a necessary condition
for (\ref{eq:delta_1}), so $G(s)$ is strictly unstable if and only
if (\ref{eq:delta_2}) holds. Substituting $\sigma$ in
(\ref{eq:approximation}) into (\ref{eq:delta_2}), we know that
$G(s)$ is strictly unstable if and only if the inequality
(\ref{eq:condition}) in Theorem \ref{th:theorem 1} holds, i.e.,
condition (c) holds if (\ref{eq:condition}) is satisfied.

So far, we have proven conditions (a), (b), and (c), and hence
guaranteed the existence of oscillatory solutions to (\ref{eq:matrix
form}). The synchronization condition for these oscillations, i.e.,
(\ref{eq:synchornization condition}),  follows easily from the
secant condition in \cite{stan_acc:07}, thus the proof is omitted
due to space limitations. Hence Theorem \ref{th:theorem 1} is
proven.
\end{proof}
\begin{Remark 1}
From (\ref{eq:synchornization condition}), we can see that with an
increase in $\gamma$, a stronger network connectivity $\varrho$ is
needed to guarantee network synchronization. Given that for $p>1$,
$\gamma$ can be verified an increasing function of the Hill
coefficient $p$, we know that a system having a higher Hill
coefficient (i.e., higher end product cooperativity) requires a
stronger coupling to maintain network synchronization.
\end{Remark 1}

\section{Oscillation period estimation based on multivariable harmonic
balance technique}
\subsection{Oscillation analysis based on  harmonic
balance technique}

In this section, we reformulate the problem of oscillation analysis
using a multivariable harmonic balance technique \cite{Iwasaki:08}.
This is motivated by the observation that $H(s)$ is a low pass
filter thus higher order harmonics of oscillations in the close-loop
system can be safely neglected. Hence the waveform of $x_{3,i}$
$(i=1,2,\ldots,N)$ can be approximated by its zero-order and
first-order harmonic components \cite{Iwasaki:08,Khalil:02}:
\begin{equation}
x_{3,i}=\alpha_i+\beta_i\sin(wt+\phi_i),\: i=1,2,\ldots,N
\end{equation}
where $\alpha_i$ and $\beta_i$ denote the amplitudes of the
zero-order and the first-order harmonic components, respectively,
and $w$ and $\phi_i$ denote the oscillation frequency and phase,
respectively.

Since $f(\bullet)$ is a static nonlinear function, it can be
approximated by its describing functions \cite{Khalil:02}:
\begin{equation}
f(x_{3,i})\approx \xi_i \alpha_i +\eta_i \beta_i\sin(wt+\phi_i)
\end{equation}
where
\begin{equation}\label{eq:describing_xi_i}
\xi_i=\frac{1}{2\pi \alpha_i}\int_{-\pi}^{\pi}f(\alpha_i +
\beta_i\sin(t))dt
\end{equation}
\begin{equation}\label{eq:describing_eta_i}
\eta_i=\frac{1}{2\pi \alpha_i}\int_{-\pi}^{\pi}f(\alpha_i +
\beta_i\sin(t))\sin(t)dt
\end{equation}
The describing function $\xi_i$ is the gain of $f(\bullet)$ when the
input is a constant  value $\alpha_i$ and the output is approximated
by its zero-order harmonic. The describing function $\eta_i$ is the
gain of $f(\bullet)$ when the input is a sinusoid of amplitude
$\beta_i$ and the output is approximated by its first-order harmonic
\cite{Khalil:02}.

Consequently, the closed-loop equations that $\alpha_i$ and
$\beta_i$ are expected to satisfy are given by \cite{Iwasaki:08}
\begin{equation}\label{eq:zero-order harmonic equation}
(I- H(0)\Xi)\vec\alpha=0
\end{equation}
and
\begin{equation}\label{eq:first-order harmonic equation}
(I- H(jw)\Pi)\vec\beta=0
\end{equation}
respectively, where
\[
\vec\alpha=\left[\begin{array}{c}\alpha_1\\
\alpha_2\\\vdots\\ \alpha_N\end{array}\right],\:
\vec\beta=\left[\begin{array}{c}\beta_1 e^{j\phi_i}\\
\beta_2e^{j\phi_2}\\\vdots\\\beta_Ne^{j\phi_N}\end{array}\right],\:
\begin{array}{c}
\Xi=\textrm{diag}\{\xi_1,\,\ldots,\xi_N\},\\\\\Pi=\textrm{diag}\{\eta_1,\,\ldots,\eta_N\}
\end{array}
\]
Therefore, the problem of oscillation analysis reduces to finding
$\vec\alpha$, $\vec\beta$, and $w$ satisfying (\ref{eq:zero-order
harmonic equation}) and (\ref{eq:first-order harmonic equation}).
Note that
 (\ref{eq:zero-order harmonic equation}) and
(\ref{eq:first-order harmonic equation}) are referred to as harmonic
balance equations.

Let $\Xi^{\ast}$ and $\Pi^{\ast}$ be constant matrices satisfying
(\ref{eq:zero-order harmonic equation}) and (\ref{eq:first-order
harmonic equation}) simultaneously. Define two linear systems
$G_0(s)$ and $G_1(s)$ as
\begin{equation}\label{eq:G_0 and G_1}
\begin{aligned}
G_{0}(s) \triangleq (I-H(s)\Xi^{\ast})^{-1}H(s),\\
G_{1}(s) \triangleq (I-H(s)\Pi^{\ast})^{-1}H(s)
\end{aligned}
\end{equation}
The systems $G_0(s)$ and $G_1(s)$ are obtained by replacing the
nonlinearity $f(\bullet)$ with the constant gain computed from the
describing functions. Thus, the two linear systems  contain some
information about the oscillations of the original nonlinear system.
According to Iwasaki \cite{Iwasaki:08}, the predicted oscillation at
frequency $\omega$ is expected stable if both $G_0(s)$ and $G_1(s)$
are marginally stable with poles of $s=0$ and $s=\pm jw$ on the
imaginary axis, respectively (the rest in the open left half plane).
Therefore, the problem of oscillation analysis can be reduced to the
following problem:

\begin{Problem 1}
For the coupled Goodwin oscillators in (\ref{eq:matrix form}), find
$\Xi^{\ast}$ and $\Pi^{\ast}$ that
\begin{itemize}
\item satisfy (\ref{eq:zero-order harmonic equation}) and
(\ref{eq:first-order harmonic equation}), respectively;
\item and at the same time guarantee that $G_0(s)$ and $G_1(s)$ in
(\ref{eq:G_0 and G_1}) are marginally stable.
\end{itemize}
\end{Problem 1}

The solution is given  in the next section.

\subsection{Oscillation period of coupled Goodwin oscillators}
According to \cite{Iwasaki:08}, (\ref{eq:zero-order harmonic
equation}) and (\ref{eq:first-order harmonic equation}) are very
difficult to solve since in general $\Xi$ and $\Pi$ depend on
$\vec\alpha$ and $\vec\beta$. Bearing in mind that we are interested
in the collective period,  we can restrict our attention to
solutions that describe synchronized oscillations of the oscillator
network. This provides a clue to solve the problem:  according to
Definition \ref{de:synchrony}, synchrony means that $x_{3,i}$ are
identical, i.e., 1) the phases  $\phi_i$ ($i=1,2,\ldots,N$) are
identical; 2) the amplitudes  $\alpha_i$ and $\beta_i$
($i=1,2,\ldots,N$) are identical. Given that $\xi_i$ and $\eta_i$
are determined by $\alpha_i$ and $\beta_i$, we further have the
equality of all $\xi_i$ and all $\eta_i$. Making use of these
properties, we have Theorem \ref{th:theorem 2}:
\begin{Theorem 2}\label{th:theorem 2}
 For the Goodwin oscillator network in (\ref{eq:matrix form}), if
 the oscillation/synchronization condition in Theorem \ref{th:theorem
 1} is satisfied, then its collective period $T_{\rm{collective}}$ is given by
\begin{equation}\label{eq:collective frequency}
T_{\rm{collective}}=\frac{2\pi}{w}=\frac{2\pi}{\sqrt{b_1b_2+b_1b_3+b_2b_3}}
\end{equation}
\end{Theorem 2}
\begin{proof}
According to the above analysis, we have
\begin{equation}\label{eq:scalarization}
{\vec\alpha}=\alpha\vec{1},\: {\vec\beta}=\beta\vec{1},\: \Xi=\xi
I_N,\:\Pi=\eta I_N
\end{equation}
where $\alpha$, $\beta$, $\xi$, and $\eta$ are constants and
$\vec{1}$ is given by:
\[
\vec{1}=\left[\begin{array}{cccc}1&1&\ldots&1\end{array}\right]^T
\]
Hence (\ref{eq:zero-order harmonic equation}) and
(\ref{eq:first-order harmonic equation}) reduce to
\begin{eqnarray}\label{eq:zero-order}
(\frac{1}{\xi}I-  H(0))\alpha\vec{1}=0
\end{eqnarray}
and
\begin{eqnarray}\label{eq:first-order}
(\frac{1}{\eta}I-  H(jw))\beta\vec{1}=0
\end{eqnarray}
respectively, which further means that $\frac{1}{\xi}$ is the
eigenvalue of $H(0)$ corresponding to the eigenvector with identical
elements, and $\frac{1}{\eta}$ is the eigenvalue of $H(jw)$
corresponding to the eigenvector with identical elements.

From (\ref{eq:H(s)}), we know the eigenvalues of $H(0)$ are
$\lambda_1=\frac{1}{b_1b_2b_3}$ and $\lambda_j
=\frac{1}{b_1(b_2+\upsilon_j)b_3}$ for $j=2,3,\ldots,N$. Further
notice that only $\lambda_1$ corresponds to eigenvectors with
identical elements. Thus we have
\begin{eqnarray}\label{eq:xi}
\frac{1}{\xi}=\lambda_1=\frac{1}{b_1b_2b_3}
\end{eqnarray}

Similarly, we can get that the eigenvalues of $H(jw)$ are
$\lambda_1=\frac{1}{(jw+b_1)(jw+b_2)(jw+b_3)}$  and $\lambda_j
=\frac{1}{(jw+b_1)(jw+b_2+\upsilon_j)(jw+b_3)}$  for
$j=2,3,\ldots,N$. Further notice that only $\lambda_1$ corresponds
to eigenvectors with identical elements. Thus we have
\begin{equation}\label{eq:eta_eigenvalue}
\frac{1}{\eta}=\lambda_1=\frac{1}{(jw+b_1)(jw+b_2)(jw+b_3)}
\end{equation}
According to (\ref{eq:describing_eta_i}), $\eta$ is real, thus the
item on the right hand side of (\ref{eq:eta_eigenvalue}) must be
real, i.e., its imaginary part is zero. Given that
\begin{equation}
\begin{aligned}
&(jw+b_1)(jw+b_2)(jw+b_3)=b_1b_2b_3-w^2(b_1+b_2+b_3) \\
&\qquad\qquad\qquad\qquad +jw((b_1b_2+b_1b_3+b_2b_3)-w^2)
\end{aligned}
\end{equation}
 we have the imaginary part $jw((b_1b_2+b_1b_3+b_2b_3)-w^2)$ equal
to $0$, which further leads to
\begin{equation}
w^2=b_1b_2+b_1b_3+b_2b_3
\end{equation}
and
\begin{equation}\label{eq:eta}
 \eta=b_1b_2b_3-(b_1b_2+b_1b_3+b_2b_3)(b_1+b_2+b_3)
\end{equation}
 Hence we know the
collective period is determined by (\ref{eq:collective frequency}).

But to prove Theorem \ref{th:theorem 2}, it remains to prove that
the oscillation at estimated frequency is stable, i.e., $G_0(s)$ and
$G_1(s)$ in (\ref{eq:G_0 and G_1}) are marginally stable
\cite{Iwasaki:08}. So we proceed to prove that: (1) $G_0(s)$ has one
pole of $s=0$ and the rest in the open left half plane and, (2)
$G_1(s)$ has a pair of imaginary poles $s=\pm jw$ and the rest in
the open left half plane.

Substituting $\Xi$ and $\Pi$ in (\ref{eq:scalarization}) into
(\ref{eq:G_0 and G_1}) yields
\begin{equation}\label{eq:G_0}
G_0(s)=(I-\xi H(s))^{-1}H(s)
\end{equation}
and
\begin{equation}\label{eq:G_1}
G_1(s)=(I-\eta H(s))^{-1}H(s)
\end{equation}
with $\xi$ and $\eta$ given in (\ref{eq:xi}) and (\ref{eq:eta}),
respectively.

We first consider $G_0(s)$ in (\ref{eq:G_0}). From (\ref{eq:G(s)}),
(\ref{eq:delta_1 in G(s)}), and (\ref{eq:delta_j in G(s)}), we know
the eigenvalues of $G_0(s)$ are given by
\[
\begin{aligned}
\delta_1 &=\frac{1}{(s+b_1)(s+b_2)(s+b_3)-\xi},\\
\delta_j &=\frac{1}{(s+b_1)(s+b_2+\upsilon_j)(s+b_3)-\xi},\quad
j=2,3,\ldots,N
\end{aligned}
\]
 Substituting $\xi$ in (\ref{eq:xi}) into the above equations, we know the poles of (\ref{eq:G_0}) are
 determined by the roots of
\begin{equation}\label{eq:roots G_0(1)}
  (s+b_1)(s+b_2)(s+b_3)-b_1b_2b_3=0
\end{equation}
and
\begin{equation}\label{eq:roots G_0(2)}
 (s+b_1)(s+b_2+\upsilon_j)(s+b_3)-b_1b_2b_3=0
\end{equation}
for $j=2,3,\ldots,N$.

It is clear that (\ref{eq:roots G_0(1)}) has one root $s=0$. And
using the Routh$-$Hurwitz stability criterion, we can get that the
other roots of (\ref{eq:roots G_0(1)}) and all roots of
(\ref{eq:roots G_0(2)}) are in the open left half plane. Hence
$G_0(s)$ is marginally stable.

Similarly, we can prove that the eigenvalues of $G_1(s)$ are
\[
\delta_1 =\frac{1}{(s+b_1)(s+b_2)(s+b_3)-\eta}
\]
  and
\[
\delta_j =\frac{1}{(s+b_1)(s+b_2+\upsilon_j)(s+b_3)-\eta}
\]
for $j=2,3,\ldots,N$, with $\eta$ given in (\ref{eq:eta}). And hence
 its poles are determined by the roots of
\begin{equation}\label{eq:roots G_1(1)}
\begin{aligned}
 & (s+b_1)(s+b_2)(s+b_3)-b_1b_2b_3\\
 & \qquad +(b_1b_2+b_1b_3+b_2b_3)(b_1+b_2+b_3)=0
\end{aligned}
\end{equation}
and
\begin{equation}\label{eq:roots G_1(2)}
\begin{aligned}
 & (s+b_1)(s+b_2+\upsilon_j)(s+b_3)-b_1b_2b_3\\
 & \qquad +(b_1b_2+b_1b_3+b_2b_3)(b_1+b_2+b_3)=0
\end{aligned}
\end{equation}
for $j=2,3,\ldots,N$.

It can be verified that $s=\pm jw$ are two roots of (\ref{eq:roots
G_1(1)}). And using the Routh$-$Hurwitz stability criterion, it
follows that the other roots of (\ref{eq:roots G_1(1)}) and all
roots of (\ref{eq:roots G_1(2)}) are in the open left half plane. So
$G_1(s)$ is marginally stable. Hence the derived oscillation at
frequency $w$ in (\ref{eq:collective frequency}) is stable, which
completes the proof.
\end{proof}
\begin{Remark 2}
The collective period in (\ref{eq:collective frequency}) is given in
terms of the dimensionless parameters in (\ref{eq:Goodwin
oscillator}). Representing the collective period with the original
dimensional parameters gives
\begin{equation}\label{eq:biological insights}
T_{\rm{collective}}=\frac{2\pi}{\sqrt{k_1k_2+k_1k_3+k_2k_3}}
\end{equation}
Eqn. (\ref{eq:biological insights}) means that the collective period
decreases with an increase in the rate constants for  degradation of
each component, but it is independent of the rates of transcription,
translation, and catalysis. These give insights into the basic
determination mechanism of the collective period  in coupled
biological oscillator networks, and may further provide guidance in
synthetic biology design.
\end{Remark 2}
\begin{Remark 3}
From (\ref{eq:biological insights}), we can see that when the
intercellular interaction is of the form in (\ref{eq:Goodwin
oscillator_entrain_network}), the collective period is only
determined by $k_1$, $k_2$, and $k_3$, and it is independent of
intercellular coupling. The results are obtained based on analytical
treatment of a network of coupled gene regulatory oscillators and
they corroborate the results in \cite{Liu:97}, which are obtained
using the phenomenological single-variable phase model and state
that the strength of intercellular coupling does not affect the
collective period of circadian rhythm oscillator networks.
\end{Remark 3}
\begin{Remark 4}
It is worth noting that if the coupling is of a form different from
(\ref{eq:Goodwin oscillator_entrain_network}), then it may affect
the collective period, even if it is still of diffusive type.
Examples have been reported in \cite{wunsche:05} and
\cite{Herrgen:10}.
\end{Remark 4}

\section{Numerical study}
In this section, simulation results are given to confirm the
theoretical predictions. We considered a network of nine Goodwin
oscillators. The coupling strengths $a_{i,j}$ were randomly chosen
from a uniform distribution on $[0,\,1]$ and are given  in Table I.
It can be verified that the coupling topology is connected and the
algebraic connectivity is $\varrho=2.4583$ ($\varrho$ is equal to
the second smallest eigenvalue of interaction matrix $A$ in
(\ref{eq:A}), as defined in Theorem \ref{th:theorem 1}).
\begin{table}[htb]
\centering
\begin{minipage}[t]{\linewidth}
\centering
 \label{ta:table1} \caption{Coupling topology $a_{i,j}\: (1\leq i,j\leq 9,\,i\neq j)$ of the Goodwin oscillator network}
\begin{tabular}{||c|c|c|c|c|c|c|c|c|c||}
\hline\hline  \backslashbox{}{}  & 1 & 2 & 3 & 4 & 5& 6& 7 & 8 & 9\\
\hline
1 &   & 0.3 & 0.5 & 0& 0.6 & 0.2 & 0 & 0.7 &0.8\\
\hline
2 & 0.3 &   & 0.7 & 0.2& 0.1 & 0.8 & 0.3 & 0.1 &0.5\\
\hline
3 & 0.5 &  0.7&   & 0.3& 0.6 & 0.2 & 0.6 & 0 &0.8\\
\hline
4 & 0 & 0.2& 0.3 &     & 0.4 & 0.6 & 0.2 & 0.9 &0.1\\
\hline
5 & 0.6 & 0.1& 0.6 & 0.4 & & 0.2 & 0.7 & 0.3 &0.8\\
\hline
6 & 0.2 & 0.8& 0.2 & 0.6 &0.2 &  & 0.1 & 0.9 &0.3\\
\hline
7 & 0 & 0.3& 0.6 & 0.2 &0.7 &0.1  & & 0.4 &0.5\\
\hline
8 & 0.7 & 0.1& 0 & 0.9 &0.3 &0.9  &0.4 &  &0.8\\
\hline
9 & 0.8 & 0.5& 0.8 & 0.1 &0.8 &0.3  &0.5 & 0.8 &\\
 \hline\hline
\end{tabular}
\end{minipage}
\end{table}

 First we tested our oscillation/synchronization condition in Theorem \ref{th:theorem
 1}.  
  The results are summarized in Table II. It can be seen that
oscillation/synchronization  can be obtained only when the
parameters satisfy $R> 1$ in (\ref{eq:condition}).
\begin{table}[htb]
\centering
\begin{minipage}[t]{\linewidth}
\centering
 \label{ta:table1} \caption{Test of the  oscillation/synchronization condition}
\begin{tabular}{||c|c|c|c|c|c||}
\hline\hline  $p$ & $b_1$ & $b_2$ & $b_3$ &
 $R$  &  Simulation results
\\
\hline 17 & 0.4 &0.4 & 0.4 & 1.7102 & Oscillation/synchronization \\
\hline 17 & 0.5 &0.5 & 0.5 & 1.6541 & Oscillation/synchronization\\
\hline 17 & 0.6 &0.6 & 0.6 & 1.5286 & Oscillation/synchronization \\
\hline 17 & 0.7 &0.7 & 0.7 & 1.3266 & Oscillation/synchronization \\
\hline 17 & 0.8 &0.8 & 0.8 & 1.0421 & Oscillation/synchronization\\
\hline 17 & 0.85 &0.85 & 0.85 & 0.8686 & No oscillation/synchronization \\
\hline 17 & 0.9 &0.9 & 0.9 & 0.676 & No oscillation/synchronization \\
\hline 17 & 1.0 &1.0 & 1.0 & 0.2620 & No oscillation/synchronization\\
\hline 17 & 0.7 &0.8 & 0.9 & 1.0433 & Oscillation/synchronization \\
\hline 17 & 0.9 &0.8 & 0.8 & 0.9300 & No oscillation/synchronization \\
\hline\hline
\end{tabular}
\end{minipage}
\end{table}

We also compared the collective periods in the oscillatory cases in
Table II with the estimated value. The results are summarized in
Table III. It can be seen that the estimated values approximate the
actual collective periods very closely.
\begin{table}[htb]
\centering
\begin{minipage}[t]{\linewidth}
\centering
 \label{ta:table1} \caption{Comparison between the estimated collective period [s] and the actual collective period [s]}
\begin{tabular}{||c|c|c|c|c|c||}
\hline\hline  $b_1=b_2=b_3$ & $0.4$ & $0.5$ & $0.6$ &
 $0.7$  &  $0.8$
\\
\hline Actual value & 10.68 & 8.00 & 6.31 & 5.23 & 4.53 \\
\hline Estimated value & 11.35 & 7.26 & 6.05 & 5.19 & 4.54 \\
\hline Estimation error & 6.27\% & -9.25\% & -4.12\% & -0.76\% & 0.22\% \\
\hline\hline
\end{tabular}
\vspace{-0.5cm}
\end{minipage}
\end{table}

\section{Conclusions}
Underlying biological rhythms are networks of interacting cellular
oscillators. How the collective oscillation patterns arise from
autonomous cellular oscillations is poorly understood. The Goodwin
oscillator is a quintessential model of biochemical oscillators
based on negative feedback mechanisms. Based on a network of coupled
Goodwin oscillators, we studied analytically the
oscillation/synchronization condition and collective period of
coupled biochemical oscillators by using a multivariable harmonic
balance technique. We give an oscillation/synchronization condition
of coupled Goodwin oscillators. The condition shows that a system
having a higher Hill coefficient (corresponding to a higher
cooperativity of end product repression) requires a stronger
intercellular coupling to maintain network synchronization. We also
analytically estimate the collective oscillation period of the
oscillator network. The collective oscillation period is only
dependent on the degradation rates of each component, and it is
independent of the rate of transcription, translation, and
catalysis. The results are confirmed by numerical simulations and
may provide guidance in synthetic-biological-oscillator design.
Given that the Goodwin oscillator has been successfully implemented
$in$ $vivo$, synthetic biology based testing of the predictions is
promising. Experimental verification is also feasible in biological
oscillators whose degradation/synthesis rates are tunable.

\bibliographystyle{unsrt}
\bibliography{abbr_bibli}

\end{document}